\documentclass[12pt]{article}

\usepackage{graphicx}
\usepackage{authblk}

\newcommand{\um}[0]{$\mu$m}

\newcommand{\mob}[0]{cm$^2$ V$^{-1}$ s$^{-1}$}

\newcommand{\sr}[0]{s_\mathrm{11}}
\newcommand{\st}[0]{s_\mathrm{21}}
\newcommand{\Iae}[0]{I_\mathrm{ae}}
\newcommand{\jae}[0]{j_\mathrm{ae}}
\newcommand{\Pin}[0]{P_\mathrm{in}}
\newcommand{\Pout}[0]{P_\mathrm{out}}
\newcommand{\GEG}[0]{\Gamma_{2\mathrm{D}}}
\newcommand{\KKeff}[0]{K^2_\mathrm{eff}}

\newcommand{\PSAW}{P_\mathrm{SAW}}
\newcommand{\pSAW}{p_\mathrm{SAW}}
\newcommand{\vSAW}{v_\mathrm{SAW}}
\newcommand{\lSAW}{\lambda_\mathrm{SAW}}
\newcommand{\xDL}{x_\mathrm{DL}}
\newcommand{\wIDT}{w_\mathrm{IDT}}

\newcommand{\dMgO}{d_\mathrm{MgO}}
\newcommand{\dZnO}{d_\mathrm{ZnO}}
\newcommand{\eMgO}{\varepsilon_\mathrm{MgO}}
\newcommand{\eZnO}{\varepsilon_\mathrm{ZnO}}

\begin{document}
\title{Acousto-electric transport in MgO/ZnO-covered graphene on SiC}

\author{Y.-T. Liou}
\author{A. Hern\'andez-M\'inguez}
\author{J. Herfort}
\author{J. M. J. Lopes}
\author{A. Tahraoui}
\author{P. V. Santos}
\affil{Paul-Drude-Institut f\"ur Festk\"orperelektronik \\ Hausvogteiplatz 5--7, 10117 Berlin, Germany
	
	yi-ting.liou@pdi-berlin.de}

\maketitle

\begin{abstract}
We investigate the acousto-electric transport induced by surface acoustic waves (SAWs) in epitaxial graphene (EG) coated by a MgO/ZnO film. The deposition of a thin MgO layer protects the EG during the sputtering of a piezoelectric ZnO film for the efficient generation of SAWs.
We demonstrate by Raman and electric measurements that the coating does not harm the EG structural and electronic properties. We report the generation of two SAW modes with frequencies around 2~GHz. For both modes, we measure acousto-electric currents in EG devices placed in the SAW propagation path. The currents increase linearly with the SAW power, reaching values up to almost two orders of magnitude higher than in previous reports for acousto-electric transport in EG on SiC. Our results agree with the predictions from the classical relaxation model of the interaction between SAWs and a two dimensional electron gas.
\end{abstract}

\vspace{2pc}

\section{Introduction}\label{sec_Intro}

Surface acoustic waves (SAWs) are elastic vibrations propagating along the surface of a solid~\cite{Rayleigh_PLMSs1-17_4_85} with typical wavelengths ranging between 0.1 and 10 \um~\cite{Datta86a, Campbell_98}. If the propagation medium is piezoelectric, then the strain wave is accompanied by a dynamic electric field with the same frequency and wavelength as the elastic vibrations~\cite{Lewis_37_85}. Both piezoelectric and strain fields have been long employed to manipulate elementary excitations in low-dimensional heterostructures placed either below the surface of a semiconductor~\cite{Rocke97a, Wiele98a, PVS152, Gell_APL93_081115_08, PVS218, Hermelin_N477_435_11, PVS223, PVS260, Schulein_PRB88_85307_13}, or on the surface of a piezoelectric insulator \cite{Kinzel_NL11_1512_11, PVS246, Regler_CP413_39_13, Weiss_NL14_2256_14, Pustiowski_APL106_13107_15, Preciado_NC6__15}. If the heterostructure contains a two-dimensional electron gas, then a SAW traversing it experiences attenuation and a shift of its propagation velocity~\cite{Wixforth_PRL56_2104_86, Wixforth89a, Simon_PRB54_13878_96}. Simultaneously, due to the acousto-electric effect~\cite{Parmenter_PR89_990_53}, a fraction of the energy lost by the SAW is transferred to the free carriers, generating electric currents in the conductive medium~\cite{Esslinger_SSC84_939_92, Falko_PRB47_9910_93, Shilton_PRB51_14770_95, Shilton_JoPCM7_7675_95, Rotter_APL73_2128_98}.

Recently, the interaction of SAWs with graphene has attracted increasing interest. Since the experimental demonstration of graphene in year 2004~\cite{Novoselov_S306_666_04}, its peculiar mechanical and electronic properties have made it a very promising candidate for applications in areas like flexible electronics, biological engineering, composite materials and even for optical electronics and photovoltaics~\cite{Ferrari_N7_4598_15}. Graphene consists of a single layer of carbon atoms forming a honeycomb lattice. Contrary to conventional two-dimensional heterostructures, this atomic configuration leads to a linear energy dispersion around the charge neutrality point. As a consequence, the charge carriers contained in graphene behave as a two-dimensional gas of relativistic Dirac fermions. The question arises whether SAWs can also be an efficient mechanism for manipulation of this kind of particles in graphene-based devices. This question has been addressed in theoretical studies about the coupling of SAWs to the graphene Dirac fermions~\cite{Thalmeier_PRB81_41409_10, Zhang_AA1_22146_11, Dietel_PRB86_115450_12, Schiefele_PRL111_237405_13}. In addition, functionalities based on the combination of SAWs and graphene have recently been demonstrated in graphene mechanically transferred to a piezoelectric substrate. Examples are the generation of SAW-induced electric currents~\cite{Miseikis_APL100_133105_12, Bandhu_APL103_133101_13, Bandhu_APL105_263106_14, Roshchupkin_JAP118_104901_15, Insepov_MA1_1495_16, Poole_SR7_1767_17}, the fabrication of SAW delay lines including graphene interdigitated transducers~\cite{Mayorov_APL104_83509_14, Emelin_APL110_264103_17}, SAW amplification by DC-voltages applied to graphene films placed on the SAW path~\cite{Insepov_APL106_23505_15}, and the demonstration of gas and light sensors based on the coupling of SAWs and graphene~\cite{Arsat_CPL467_344_09, Chivukula_ITUF59_265_12, Whitehead_APL103_63110_13, Poole_APL106_133107_15}.

For future commercial applications, however, it is desirable the demonstration of these functionalities in material combinations that make possible large scale fabrication at relatively low cost. From this point of view, epitaxial graphene (EG) on SiC is a promising candidate, because it allows the formation of large area graphene layers by Si sublimation from the SiC surface~\cite{Berger_JPCB108_19912_04, Berger_S312_1191_06}. Furthermore, as the EG layers are prepared directly on an insulating substrate, they can be processed straight away into devices using conventional planar fabrication techniques. Due to the weak piezoelectricity of SiC, a strong piezoelectric film must be included for the efficient generation of SAWs and to enhance the acousto-electric coupling between the SAW and the electron gas in graphene. Recently, we have demonstrated acousto-electric currents in EG on SiC~\cite{PVS269, PVS283} using ZnO as piezoelectric layer. To avoid the damage of the graphene during the sputtering of ZnO, an hydrogen-silsesquioxane (HSQ) interlayer was placed on top of the EG prior to ZnO deposition. Although the HSQ coating fulfilled its protective role, we also observed a significant decrease in the EG carrier mobility~\cite{PVS283}. This is deleterious for devices based on the acousto-electric interaction, because the intensity of the acousto-electric current depends directly on the mobility of the free charge carriers~\cite{Rotter_APL73_2128_98}. Therefore, the use of protective layers that also preserve the graphene electronic properties is a key issue towards the efficient exploitation of acousto-electric devices in EG on SiC.

In this contribution, we report on the generation of acousto-electric currents in ZnO-coated EG using a thin film of MgO as protective layer. MgO has been identified as a promising insulator in graphene-based devices, e.g. as efficient tunnel barrier between graphene and ferromagnetic layers for spin injection~\cite{Han2010, Volmer2013, Volmer2014, Droegeler2014}. In addition, the growth of atomically smooth MgO on graphene by molecular beam epitaxy has also been demonstrated~\cite{Wang_APL93_183107_08, Godel2013}. We confirm by Raman spectroscopy and electric characterization that both the structural and electronic properties of the EG are preserved after deposition of the MgO and ZnO top layers. The acoustic response of the interdigital transducers (IDTs) placed on the ZnO demonstrates the generation of two SAW modes with different propagation velocities. We show that both SAW modes induce acousto-electric currents in EG structures patterned in the SAW propagation path, and that the current densities are about one order of magnitude larger than the best values measured in our previous devices using HSQ as protective layer~\cite{PVS283}. Finally, the dependence of the acousto-electric current on the applied SAW power agrees well with the behavior expected from the relaxation model of the acousto-electric effect.

We have organized the manuscript as follows. Section~\ref{sec_Sample} describes the fabrication of the acousto-electric devices. Section~\ref{sec_Resul} presents the structural, electrical and acoustic characterization of our samples, as well as the acousto-electric currents measured in the EG. The intensity of the SAW-induced currents is discussed in Section~\ref{sec_Disc} in the frame of the relaxation model of the interaction between SAWs and a two dimensional electron gas. We conclude the manuscript by summarizing our results in Section~\ref{sec_Concl}.

\section{Sample fabrication}\label{sec_Sample}

\begin{figure}
\centering
\includegraphics[width=\linewidth]{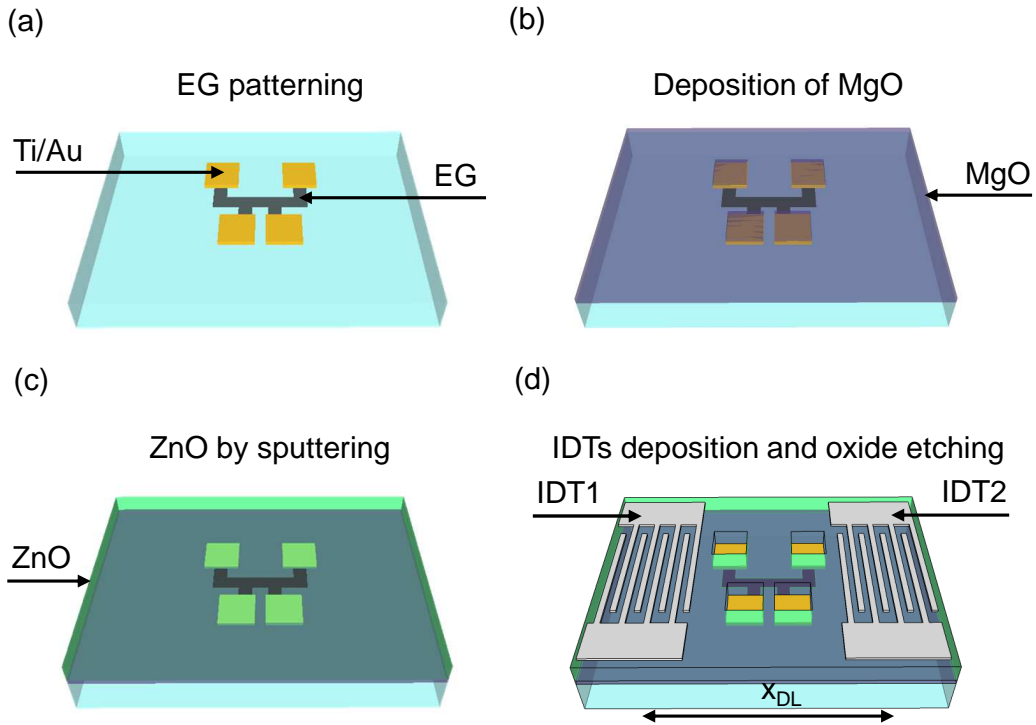}
\caption{Steps of sample fabrication. (a) Selective etching of graphene (dark grey stripes) on SiC (light blue) and contact metalization (yellow squares). (b) Deposition of poly-crystalline MgO film (dark blue) in ultra-high vacuum chamber. (c) Sputtering of piezoelectric ZnO (green) layer on top of MgO. (d) Patterning of interdigital transducers (IDTs, light grey) by optical lithography, and etching of the MgO/ZnO bilayer on the graphene electric contacts. The figures are not to scale.}
\label{fig_processing}
\end{figure}

\begin{figure}
	\centering
	\includegraphics[width=.8\linewidth]{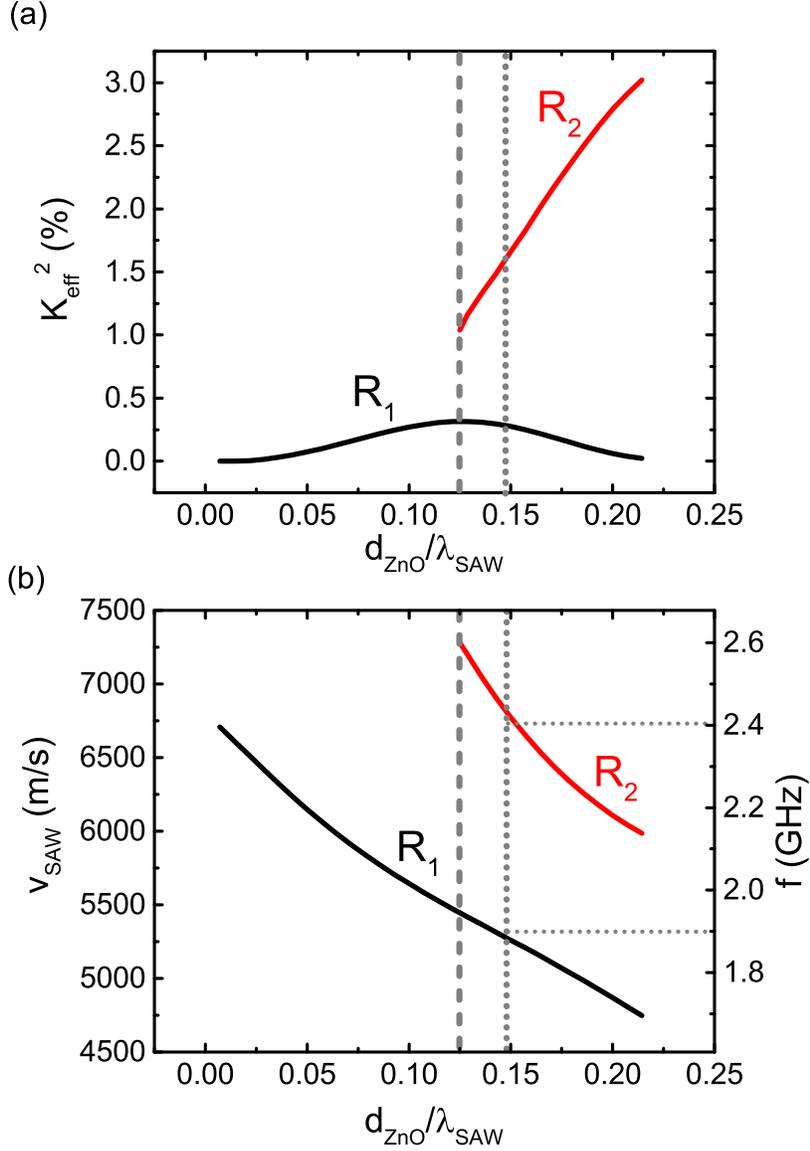}
	\caption{(a) Numerically simulated acousto-electric coupling coefficient, $\KKeff$, as a function of the ratio between the ZnO thickness, $\dZnO$, and SAW wavelength, $\lSAW$. In the simulation, the thickness of the MgO layer and SAW wavelength were kept constant at $\dMgO=15$~nm and $\lSAW=2.8~\mu$m, respectively. The black curve corresponds to the first Rayleigh mode, $R_1$. The red curve represents the results for the second Rayleigh mode, $R_2$, which appears for ratios $\dZnO/\lSAW\geq0.125$. (b) SAW phase velocity, $\vSAW$, as a function of $\dZnO/\lSAW$. The right scale displays the SAW frequency $f=\vSAW/\lSAW$. In both panels, the dashed line indicates the nominal $\dZnO/\lSAW$ ratio used in our device, while the dotted line corresponds to the value at which the calculated SAW frequencies better agree with the measured ones.}
	\label{fig_k2_and_v}
\end{figure}

We show in Fig.~\ref{fig_processing} a sketch of the fabrication steps of our devices.  First, monolayer EG was synthesized on a $10\times10$~mm$^2$ 4H-SiC(0001) substrate by silicon sublimation~\cite{Berger_S312_1191_06, Emtsev_NM8_203_09, Oliveira_APL99_111901_11}. To characterize the electronic properties of the EG layer, we performed Hall resistance measurements at room temperature using the van der Pauw method. For the sample discussed here, we obtained $\mu=825\pm1$~\mob~and $n=3.490\times10^{12}\pm 0.002\times10^{12} $~cm$^{-2}$ for the mobility and carrier density, respectively. The sample was then patterned into stripes of $w=10$~\um~width and several lengths by oxygen plasma etching, followed by the evaporation of Ti/Au pads for electric contact, cf. Fig.~\ref{fig_processing}.1. Next, the sample was placed in an ultra-high vacuum chamber, where we deposited a poly-crystalline MgO layer of $\dMgO=15$~nm thickness. This was done at a substrate temperature of 350$^{\circ}$C by sublimation of pure Mg and providing molecular oxygen, hence allowing for a precise control of stoichiometry and growth ratio. As MgO is a non-piezoelectric material, we selected a thickness that was thin enough to assure the strong coupling of the top piezoelectric ZnO layer with the graphene free charges. In addition, it was thick enough to protect the graphene structures during the sputtering of the $\dZnO=350$~nm-thick ZnO film, cf. Fig.~\ref{fig_processing}.3. The temperature and gas pressure conditions of the sputtering process ensured that the c-axis of the ZnO crystallites are predominantly oriented perpendicular to the sample surface, which guarantees the piezoelectricity required for the efficient SAW generation. As can be seen from our numerical simulations displayed in Fig.~\ref{fig_k2_and_v}(a), the thickness of the ZnO layer was selected in order to obtain the largest acousto-electric coupling coefficient, $\KKeff$, of the fundamental Rayleigh mode, $R_1$, for the SAW wavelength used in our experiments, $\lSAW=2.8~\mu$m. In addition, this thickness also allows the generation of a second, high velocity Rayleigh mode, $R_2$. This behavior is characteristic of multilayer structures, where the acoustic velocity of the top film is much lower than that of the substrate, like ZnO on SiC~\cite{Didenko_IToUFaFC47_179_00}.

The SAWs were generated by IDTs deposited on the ZnO layer by photo-lithography and metal evaporation, cf. Fig.~\ref{fig_processing}.4. Each IDT consists of 150 finger pairs with $\lSAW=2.8$~\um~periodicity, 50\% metalization ratio and an aperture of $\wIDT=50$~\um. The length of the SAW delay lines, $\xDL$, defined as the distance between the centers of each pair of IDTs (c.f. Fig.~\ref{fig_processing}), ranges from 1500~\um~to 1700~\um. Finally, selective etching of the MgO/ZnO layer on the gold pads areas allowed the electric contact to the EG stripes.

\section{Experimental Results}\label{sec_Resul}

\subsection{Graphene characterization}

\begin{figure}
\centering
\includegraphics[width=0.8\linewidth]{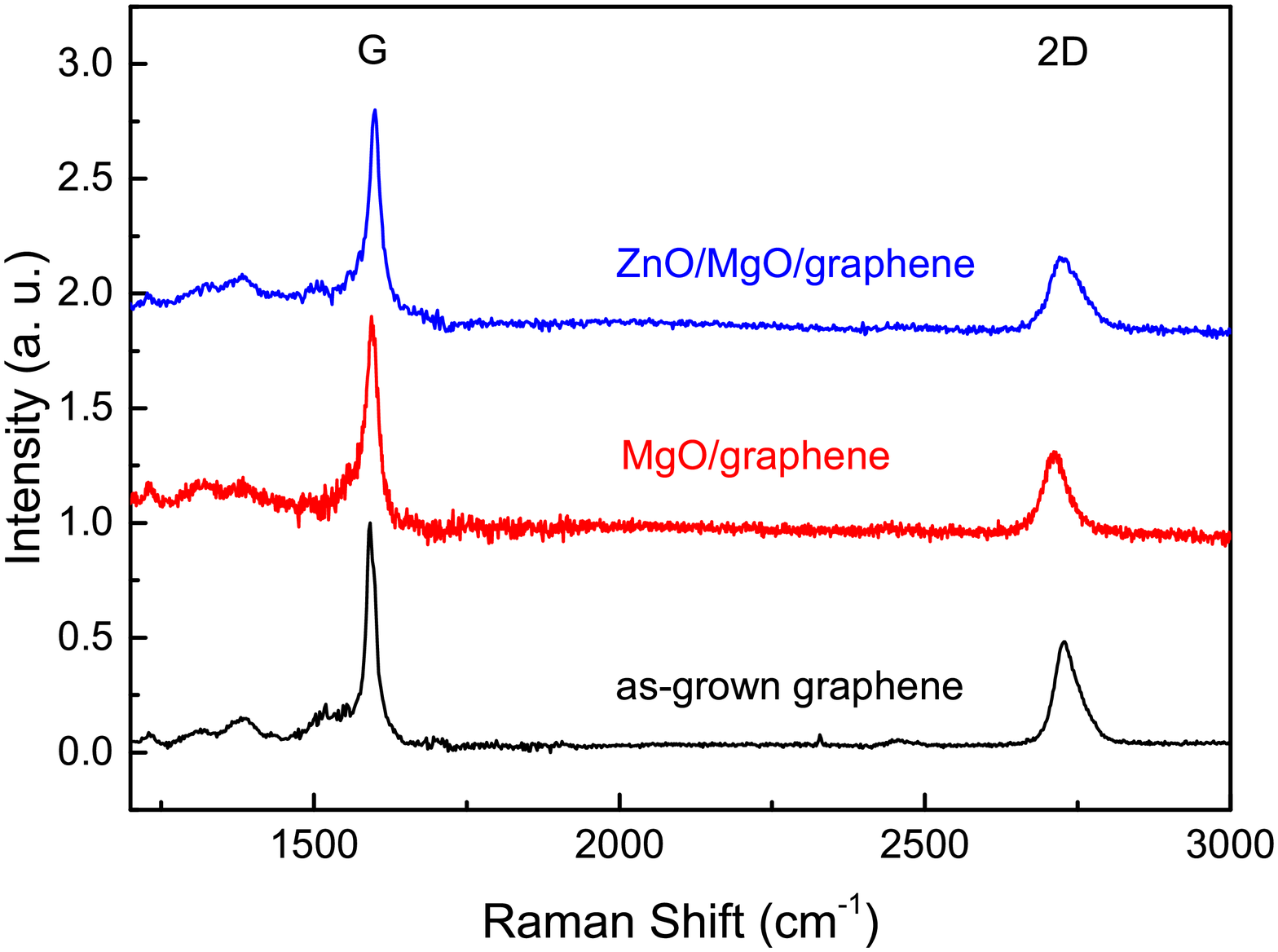}
\caption{Raman characterization of pristine EG on SiC (black curve), as well as after deposition of MgO (red curve), and ZnO (blue curve). The contribution of the SiC substrate to the Raman signal has been subtracted in all cases. The curves are normalized and vertically shifted for clarity.}
\label{fig_Raman}
\end{figure}

We used Raman spectroscopy to investigate the structural quality of our EG after the deposition of the MgO/ZnO bilayer. The black curve in Fig.~\ref{fig_Raman} displays the Raman spectrum of as-grown graphene obtained using a 473~nm laser source. It shows the expected G and 2D peaks at 1593~cm$^{-1}$ and 2730~cm$^{-1}$, respectively. The red curve displays the Raman spectrum after coating the EG with the MgO film, while the blue curve corresponds to the measurement taken after the sputtering of the ZnO layer. Apart from a slight blue shift of the G line of about $4.3\pm3.0$~cm$^{-1}$, there are no significant differences between the spectra. The D peak around 1365~cm$^{-1}$, which provides information about the presence of defects in the EG layer~\cite{Malard_PR473_51_09}, has a very low intensity, and thus remains in all cases hidden in the multi-line trace characteristic of the buffer layer that typically forms between the EG and the SiC substrate~\cite{Fromm_NJP15_43031_13}. In addition, mappings of the G and 2D lines across the EG stripes confirm that, after the sputtering process, the graphene is still continuous, without the presence of holes and cracks. This demonstrates that (\textit{i}) the deposition of MgO did not degrade the EG quality, and (\textit{ii}) the MgO layer protected the underlying graphene from being damaged during the sputtering of ZnO. 

\begin{figure}
\centering
\includegraphics[width=0.9\linewidth]{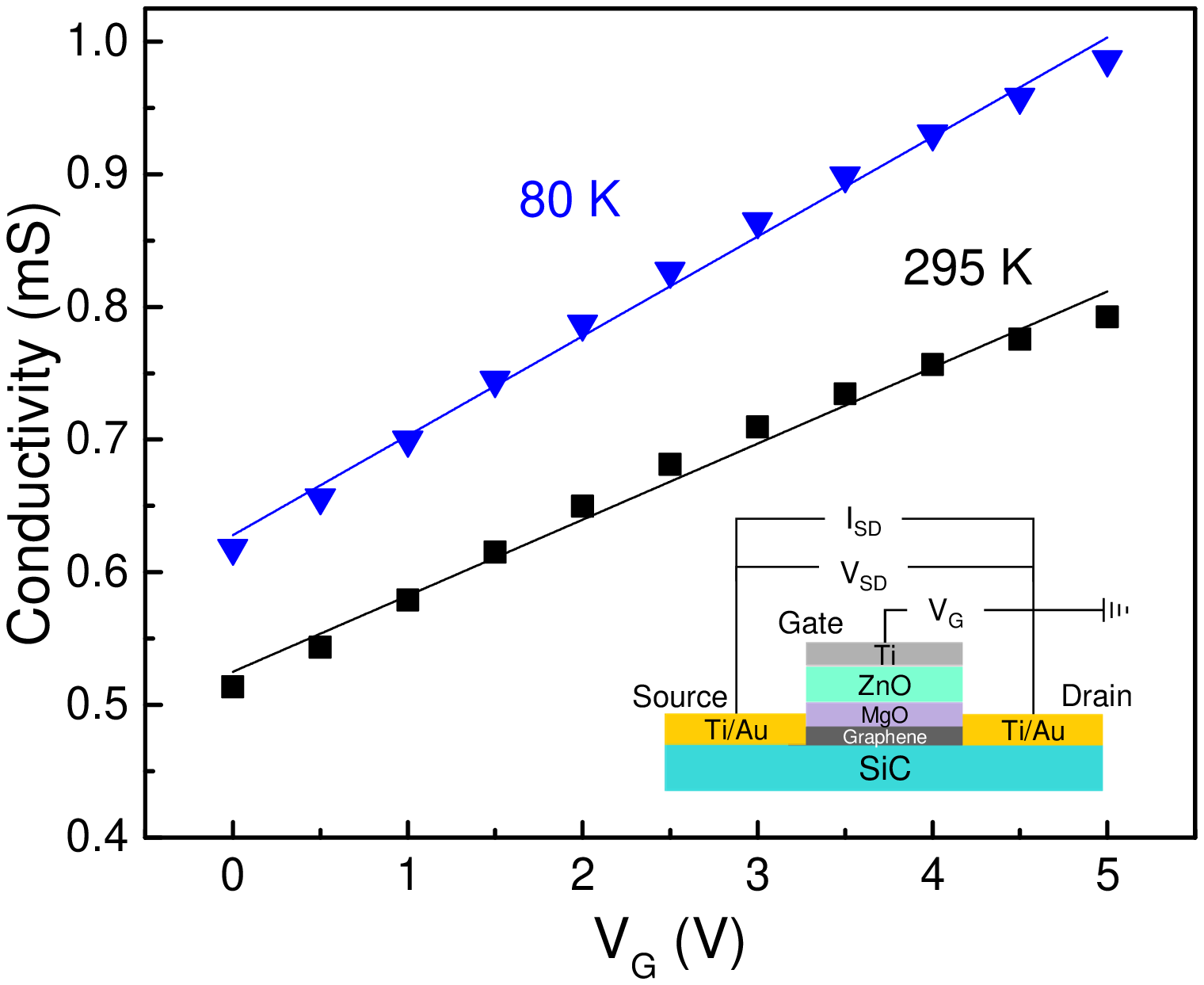}
\caption{Conductivity of EG measured at room temperature (black squares) and at 80 K (blue triangles) as a function of the gate voltage, $V_G$, applied to the field-effect-transistor displayed in the inset. The lines are linear fits to the data.}
\label{fig_FET}
\end{figure}

We probed the electronic properties of our coated graphene by measuring the performance of field-effect transistors (FETs) patterned on the sample. The FETs consist of 14~\um-long graphene stripes acting as the electronic channel between source (S) and drain (D). The MgO/ZnO layers serve as the dielectric medium for the application of the top gate (G) voltage  (see inset of Fig.~\ref{fig_FET}) via a 10~nm-thick titanium layer evaporated on top of the ZnO areas covering the graphene stripes. Figure~\ref{fig_FET} shows the dependence of the graphene sheet conductivity, $\sigma$, on the voltage bias applied to the top gate, $V_G$. It was estimated by measuring the source-drain current, $I_{SD}$, as a function of the source-drain voltage, $V_{SD}$, for each value of $V_G$ at both room temperature (black squares) and at 80~K (blue triangles). The linear dependence of $\sigma$ within the applied range of gate voltages agrees well with the Drude prediction $\sigma=ne\mu$. Here, $\mu$ is the EG carrier mobility, $e$ is the electron charge, and $n=(V_G-V_0)\times C/e$ is the electronic carrier density. The latter is proportional to the gate capacitance per unit area, $C$, and to the difference between the applied gate voltage and the value at which the graphene reaches the carrier neutrality point, $V_0$. The increase of $\sigma$ for $V_G\geq0$ indicates that graphene is $n$-doped, as expected for monolayer graphene on the Si face of SiC~\cite{First_MB35_296_10}.

From the slope of $\sigma(V_G)$, we estimated the field effect mobility of our MgO/ZnO capped graphene devices according to:

\begin{equation}\label{eq_mobility}
\mu=C^{-1}\frac{d\sigma}{dV_G},
\end{equation}

\noindent where $C=19$ nF/cm$^2$ was calculated from the thickness of the dielectric layers as:

\begin{equation}
C=\varepsilon_0\left[\frac{\dMgO}{\eMgO}+\frac{\dZnO}{\eZnO}\right]^{-1}.
\end{equation}

\noindent Here, $\varepsilon_0$, $\eMgO=9.8$ and $\eZnO=7.9$ are the vacuum permittivity and the relative permittivities of the MgO and ZnO layers, respectively~\cite{Hanada2009}. We obtain a carrier mobility of $\mu=2970\pm120$~\mob~at room temperature, and $3890\pm100$~\mob~at 80~K. These values demonstrate that the presence of the MgO layer does not deteriorate the electronic properties of the EG. The enhancement of $\mu$ as the temperature decreases is consistent with the main mechanism limiting the carrier mobility of EG on SiC at room temperature, namely electron-phonon scattering mediated by the buffer layer~\cite{Jobst_PRB81_195434_10, Speck_APL99_122106_11}. From $\sigma$ at $V_G=0$ and the calculated carrier mobility, we estimate a carrier density $n=\sigma/(e\mu)\approx1.2\times10^{12}\pm0.1\times10^{12}$~cm$^{-2}$. The lower carrier density and higher mobility with respect to the values obtained in the pristine EG could suggest a certain degree of hole doping by the MgO layer.

\subsection{SAW generation and transmission}

\begin{figure}
\centering
\includegraphics[width=\linewidth]{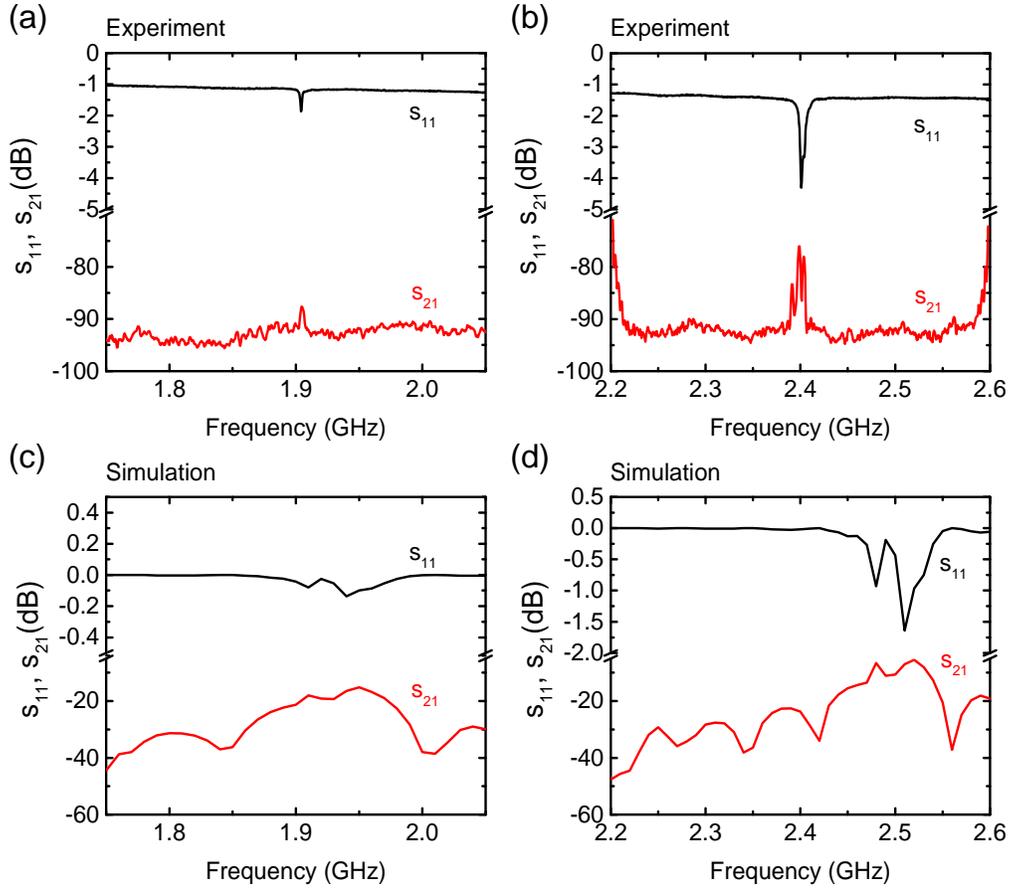}
\caption{Measured power reflection and transmission coefficients, $\sr$ and $\st$, respectively, for the two SAW modes generated by the IDT: (a) $f_a=1.9$~GHz, and (b) $f_b=2.4$~GHz. Panels (c) and (d) show the corresponding numerical simulations. Note the different vertical scales for the $\sr$ coefficients.}
\label{fig_Scoeff}
\end{figure}

To study the SAW generation and propagation efficiency in our multilayered structure, we have measured the rf-power reflection and transmission scattering coefficients, $\sr$ and $\st$, for the delay line in Fig.~\ref{fig_processing}(d), consisting of IDT$_1$ and IDT$_2$. The black curves of Figs.~\ref{fig_Scoeff}(a) and \ref{fig_Scoeff}(b) display the magnitude of the frequency resolved $\sr$ spectra for the two SAW Rayleigh modes generated by the IDTs, with frequencies $f_a=1.904$~GHz and $f_b=2.401$~GHz. This is in agreement with our simulations of SAW velocities and frequencies displayed in Fig.~\ref{fig_k2_and_v}(b) for the multilayer structure. Both modes correspond to SAWs with the wavelength of the IDT, $\lSAW=2.8$~\um, but different phase velocities $v_a=\lSAW f_a=5331$~m/s and $v_b=\lSAW f_b=6723$~m/s. The slight discrepancy between simulated and experimental values are probably related to small differences in the nominal layer thicknesses and acoustic properties used in the simulations with respect to the real ones.

The observed acoustic modes can be reproduced by the numerical simulation of the $s$ parameters, cf. Figs.~\ref{fig_Scoeff}(c) and \ref{fig_Scoeff}(d). These were obtained by using a finite element approach~\cite{getdp,gmsh_general} that simulates a two-dimensional $x-z$ cross-section of the delay line in Fig.~\ref{fig_processing}(d) ($x$ corresponds to the SAW propagation direction, and $z$ to the sample depth). In our calculations, we used the same layer thickness, finger periodicity and metalization ratio of the IDTs as in the real sample. To make the problem numerically tractable, we reduced the number of finger pairs to 20 and the length of the SAW delay line to just 25 $\lSAW$. Our simulations reproduce well the two frequency modes, and the fact that the acousto-electric efficiency of the high frequency mode is stronger than that of the low frequency one (this is reflected in the larger $\sr$ dip of $f_b$ with respect to the one of $f_a$). The wider resonance width observed in the calculations is due to the smaller number of finger pairs with respect to the real case~\cite{Campbell_98}. We also observe a small blue-shift of the calculated resonances with respect to the experimental ones, which we attribute to slight discrepancies in the nominal thicknesses and material properties used in the calculation with respect to the real ones.

Figure~\ref{fig_Scoeff} also displays the experimental and simulated $\st$ spectra of the two SAW modes (red curves). The experimental data were acquired using time-gate filtering to reject the contribution of the rf cross-talk between the IDTs. We observe that the amplitude of the measured transmission peaks is much weaker than that of the simulated data. Although a certain degree of SAW attenuation is expected due to the acousto-electric coupling between SAWs and graphene stripes patterned along the delay line, we show in Section~\ref{sec_Disc} that it is not strong enough to account for the observed results.
We attribute the low intensity of the $\st$ spectra to (\textit{i}) differences in the resonant frequency and acousto-electric efficiency of the two IDTs in the delay line, and (\textit{ii}) SAW attenuation due to dispersion and/or absorption along the propagation path. To quantify this effect, we proceed as follows: the SAW power leaving IDT$_1$ towards IDT$_2$, $\PSAW$, results from the acousto-electric conversion efficiency, $\alpha_1$, of the nominal rf-power applied to IDT$_1$, $\Pin$, which can be estimated for each SAW mode by applying the following expression to Figs.~\ref{fig_Scoeff}(a) and \ref{fig_Scoeff}(b):

\begin{equation}\label{eq_SAWpower}
\alpha_1=\frac{\PSAW}{\Pin}=\frac{1}{2}\left[10^{\sr^{(non-res)}/10}-10^{\sr^{(res)}/10}\right].
\end{equation}

\noindent Here, $\sr^{(res)}$ and $\sr^{(non-res)}$ are the $\sr$ coefficients measured at the resonance frequency and away from it, respectively. The factor 1/2 accounts for the fact that the IDT launches two acoustic beams in opposite directions. In an ideal acoustic delay line, the SAW power leaving IDT$_1$ arrives to IDT$_2$ without losses, where a fraction is transformed back into rf-power, $\Pout=\alpha_2\PSAW$. As an IDT is a passive device, the conversion efficiency $\alpha_2$ from SAW to rf is the same as from rf to SAW~\cite{Campbell_98}, and therefore it can be calculated for IDT$_2$ applying Eq.~\ref{eq_SAWpower} to $s_{22}$. Therefore, $\st^\star=10\log(\Pout/\Pin)=10\log(\alpha_1\alpha_2)$ is the expected transmission coefficient for the delay line in the absence of SAW attenuation. For the case discussed in Fig.~\ref{fig_Scoeff}, it gives $\st^\star(f_a)\approx-41$~dB and $\st^\star(f_b)\approx-26$~dB. The values acquired by our network analyzer, however, are $\st(f_a)=-88$~dB  and $\st(f_b)=-76$~dB. We attribute this difference to SAW attenuation while it travels along the delay line. We account for this attenuation by assuming that $\st=10\log[\alpha_1\exp\left(-\xi\xDL\right)\alpha_2]$, where $\xi$ is the SAW attenuation rate during propagation. Using these suppositions, we obtain $\xi_a=7.2\times10^{-3}$~\um$^{-1}$ and $\xi_b=7.6\times10^{-3}$~\um$^{-1}$ for each acoustic mode.

\subsection{SAW-induced electric current}\label{subsec_Iae}

\begin{figure}
\centering
\includegraphics[width=0.7\linewidth]{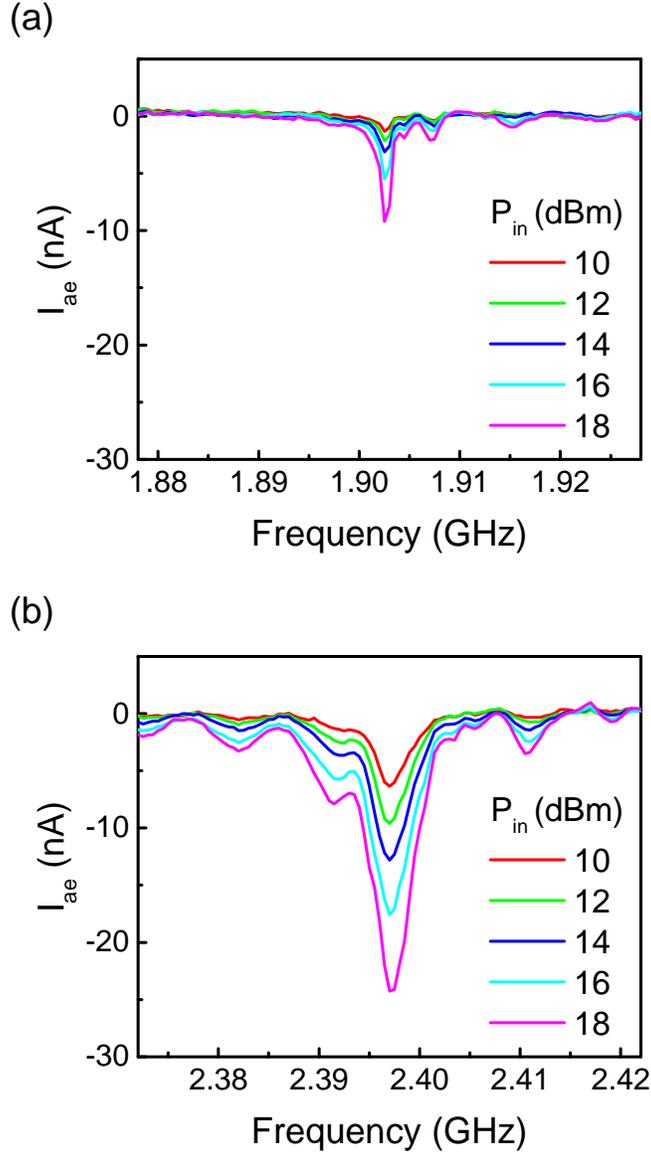}
\caption{Acousto-electric current, $\Iae$, measured in a graphene stripe as a function of the rf-frequency applied to the IDT. The experiment was performed for frequencies around the SAW modes (a) $f_a=1.9$~GHz, and (b) $f_b=2.4$~GHz, and for several nominal input rf-powers, $\Pin$.}\label{fig_AEcurrents}
\end{figure}

We have measured the acousto-electric current, $\Iae$, induced by the SAWs in four graphene stripes, each of them placed at the acoustic path of a different SAW delay line in the sample. For each graphene stripe, we recorded the electric current with a Keithley 2602 multimeter while scanning the frequency, $f$, and power, $\Pin$, of the rf-signal applied to one of the IDTs of the corresponding delay line. Figure~\ref{fig_AEcurrents}(a) displays $\Iae$ for one of the tested stripes, measured over the frequency range of $f_a$. For each value of $\Pin$, there is a clear current peak when $f$ coincides with the resonant frequency of the IDT. In addition, the amplitude of this current peak increases with $\Pin$, reaching $\Iae\approx10$~nA for the largest rf-power used in our experiment. This corresponds to a linear current density $\jae=\Iae/w=10^{-3}$~A/m, which is one order of magnitude larger than the values achieved in our previous samples, where we had used the same IDT design and nominal thickness of the ZnO film, but HSQ as protective layer~\cite{PVS283}. We attribute this enhancement of $\Iae$ under similar frequency and power of the rf applied to the IDT to the better electronic properties of EG coated by MgO with respect to the one coated by HSQ: while the carrier mobility in our current device is about $3000$~\mob, it was only $100$~\mob~in the devices using HSQ.

Figure~\ref{fig_AEcurrents}(b) displays the results obtained for the same values of $\Pin$ as in panel (a), but now scanning the rf-frequency around $f_b$. For each value of $\Pin$, the amplitude of the acousto-electric current measured at $f_b$ is always larger than the acquired at $f_a$, obtaining $\Iae\approx25$~nA for the largest $\Pin$ applied. We attribute this to the fact that, as already discussed in Fig.~\ref{fig_Scoeff}, the SAW generation efficiency of the IDTs in our SiC/MgO/ZnO multilayer is better for the $f_b$ than for the $f_a$ mode. Again, this result is an improvement with respect to the previous devices using HSQ, where no acousto-electric current was observed for this SAW mode. 

\section{Discussion}\label{sec_Disc}

\begin{figure}
\centering
\includegraphics[width=0.7\linewidth]{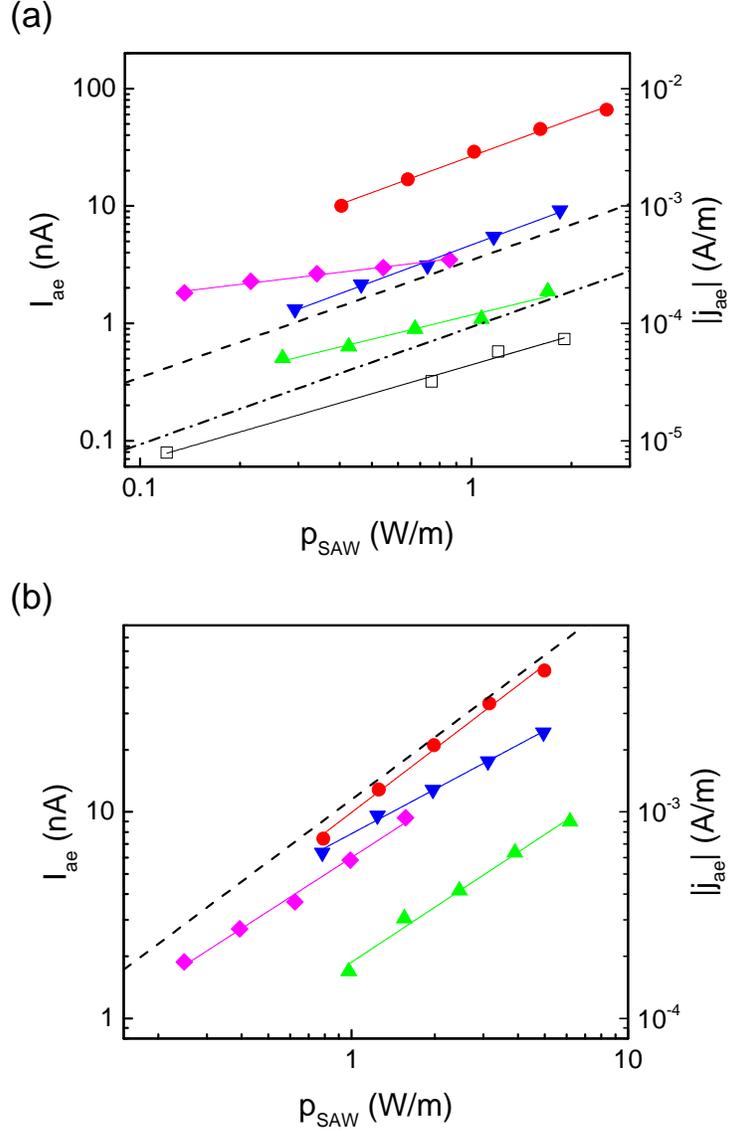}
\caption{Amplitude of the acousto-electric current, $\Iae$, and its corresponding current density, $\jae$, as a function of the SAW linear power density, $\pSAW$, for the acoustic modes (a) $f_a=1.9$~GHz, and (b) $f_b=2.4$~GHz. The symbols correspond to devices A (red circles), B (green up triangles), C (blue down triangles), and D (violet diamonds). The open black squares are from Ref.~\cite{PVS283}. The solid lines are fits to the data. The dashed and dot-dashed lines correspond to the theoretical predictions according to Eq.~\ref{eq_jae} for the present devices and for the one from Ref.~\cite{PVS283}, respectively.}\label{fig_AEsummary}
\end{figure}

According to the relaxation model of the acousto-electric effect~\cite{Shilton_PRB51_14770_95, Rotter_APL73_2128_98, Bandhu_APL103_133101_13}, the interaction between a SAW and a two-dimensional electron gas placed close to the surface induces an acousto-electric current density, $\jae$, which can be expressed as:

\begin{equation}\label{eq_jae}
\jae= -\mu \frac{\pSAW}{\vSAW}\GEG.
\end{equation}

\noindent Here, $\mu$ is the carrier mobility, $\vSAW$ the SAW velocity, $\pSAW=\PSAW/\wIDT$ the linear SAW power density, and $\GEG$ is the SAW attenuation rate due to the interaction of the SAW with the charge carriers of the two-dimensional gas.  This model has been successfully applied to SAW-induced acousto-electric currents in graphene mechanically transferred to the surface of a LiNbO$_3$ substrate~\cite{Miseikis_APL100_133105_12, Bandhu_APL103_133101_13, Bandhu_APL105_263106_14, Poole_SR7_1767_17}. In our experiment, the EG is not at the top of the piezoelectric layer, but at a depth $d=\dZnO+\dMgO=365$~nm with respect to the sample surface. However, as $d/\lSAW=0.13\ll 1$, most of the SAW still extends into the region below the EG, and thus we expect the model to be also a good approximation for our MgO/ZnO coated graphene on SiC.

To test the validity of Eq.~\ref{eq_jae}, we take into account that $\GEG$ depends on the graphene electric conductivity $\sigma$ according to~\cite{Wixforth89a}:

\begin{equation}\label{eq_Gamma}
\GEG=\KKeff\frac{\pi}{\lSAW}\frac{\sigma/\sigma_M}{1+\left(\sigma/\sigma_M\right)^2},
\end{equation}

\noindent where $\sigma_M=\vSAW\varepsilon_0\left(\varepsilon_{SiC}+\varepsilon_{MgO}\right)$ is a characteristic conductivity that depends on the SAW velocity and the dielectric constants of the layers directly below and above the graphene layer, $\varepsilon_{SiC}\approx\varepsilon_{MgO}=9.6$ in this case. In our sample, $\sigma/\sigma_M\sim 500$ for both Rayleigh modes. Therefore, $\GEG\propto\sigma_M/\sigma$ and Eq.~\ref{eq_jae} can be approximated as:

\begin{equation}\label{eq_jae2}
\jae\approx-\mu\pi \KKeff\frac{\varepsilon_0\left(\varepsilon_{SiC}+\varepsilon_{MgO}\right)}{\lSAW\sigma}\pSAW. 
\end{equation}

\noindent This means that, in our experimental regime, the amplitude of the acousto-electric current depends essentially on $\KKeff$ and the SAW power generated by the IDT, but not on the SAW velocity.

We have summarized in the log-log scale of Fig.~\ref{fig_AEsummary}(a) the experimental values of $\Iae$, and their corresponding current density, $\jae$, measured at $f_a$ for the four tested graphene devices as a function of the SAW power density, $\pSAW$, acting on the device. The latter was estimated as $\pSAW=\alpha\Pin\exp(-\xi_a x_0)/\wIDT$, where $\alpha$ is the electro-acoustic conversion efficiency of the IDT used in each case, and $\exp(-\xi_a x_0)$ accounts for the SAW attenuation along the distance between the IDT output and the center of the tested graphene stripe, $x_0$. Figure~\ref{fig_AEsummary}(b) shows the corresponding results for the case of $f_b$. Except for mode $f_a$ of device D, the slope of the linear fits to the data (solid lines) approaches unity in all cases, thus confirming the linear dependence of $\jae$ with respect to $\pSAW$ predicted by Eq.~\ref{eq_jae}. For both SAW modes, we also observe a spread in the values of $\jae$ measured in different graphene devices, which we attribute to fluctuations in their electronic properties. However, in spite of these fluctuations in the electronic quality, the acousto-electric current densities are always larger than the ones reported in our previous device using HSQ as protective layer (displayed in Fig.~\ref{fig_AEsummary}(a) as open squares), thus confirming the better quality of the EG coated by MgO. 

We have also calculated the values of $\jae$ expected from Eq.~\ref{eq_jae2} for each SAW mode. To do this, we used the graphene carrier mobility and electric conductivity at room temperature from Eq.~\ref{eq_mobility} and Fig.~\ref{fig_FET}, respectively. We estimated $\KKeff$ by numerically solving the coupled mechanical and electromagnetic differential equations of a $\lambda=2.8$~\um~SAW propagating along our multilayer, and comparing the velocity difference between SAWs propagating along a short-circuited or open-circuited top surface~\cite{Campbell_98}. The results are $K^2_{eff}(f_a)=0.314$~\% and $K^2_{eff}(f_b)=1.04$~\%, cf. Fig.~\ref{fig_k2_and_v}(a), in qualitative agreement with the experimental data of Fig.~\ref{fig_Scoeff}, where the generation efficiency of the $f_b$ mode is stronger than the $f_a$ mode. We have displayed the theoretical prediction of $\jae$ as dashed lines in Fig.~\ref{fig_AEsummary}, obtaining a reasonable agreement in the order of magnitude with the experimental results.

Using Eq.~\ref{eq_Gamma}, we have estimated $\GEG$ and compared it with the attenuation rates $\xi_a$ and $\xi_b$ obtained in Section~\ref{subsec_Iae}. We obtain $\GEG^{(a)}=6.4\times10^{-6}$~\um$^{-1}$ and $\GEG^{(b)}=27\times10^{-6}$~\um$^{-1}$ for SAW modes $f_a$ and $f_b$, respectively. Taking into account the length and number of graphene stripes placed along the delay line, the attenuation due to the acousto-electric effect is expected to be less than 0.05~dB, far too low to account for the attenuation values discussed in Section~\ref{subsec_Iae}. The mechanisms responsible for SAW attenuation in our sample must therefore be independent of the presence of EG. Its origin could be related e.g. to a stronger acoustic scattering by the MgO and ZnO crystallites than originally expected. A better understanding of this will require additional SAW characterization e.g., interferometric mapping of the SAW field~\cite{PVS281}, which goes beyond the scope of this manuscript. 

\begin{figure}
\centering
\includegraphics[width=0.7\linewidth]{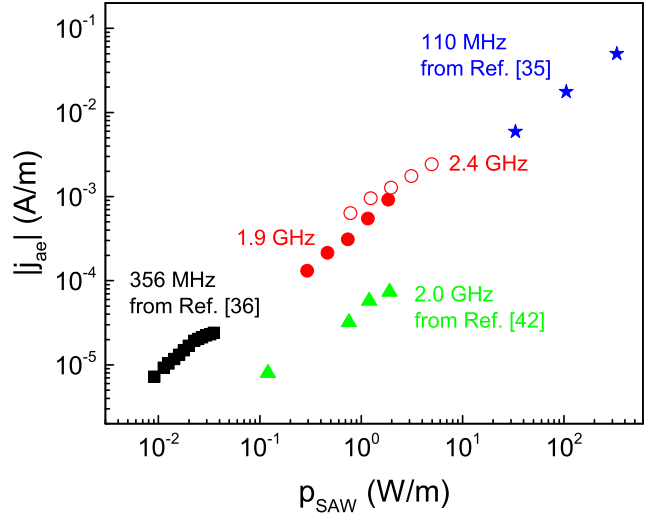}
\caption{Acousto-electric current density, $\jae$, as a function of the SAW power density, $\pSAW$, for the two SAW modes of the device in Fig.~\ref{fig_AEcurrents} (red solid and open circles for $f_a$=1.9~GHz and $f_b=2.4$~GHz, respectively), and for graphene transferred to LiNbO$_3$ estimated from Ref.~\cite{Miseikis_APL100_133105_12} (blue stars) and Ref.~\cite{Bandhu_APL103_133101_13} (black squares). The green triangles correspond to the values reported in our devices using HSQ as protective layer~\cite{PVS283}.}\label{fig_AEcomparison}
\end{figure}

Finally, we have compared the acousto-electric performance of our devices based on epitaxial graphene on SiC with the ones based on CVD graphene transferred to the surface of a strong piezoelectric substrate like 128$^\circ$ Y-cut LiNbO$_3$. To do this, we have displayed in Fig.~\ref{fig_AEcomparison} the amplitude of $\jae$ as a function of $\pSAW$ for our device shown in Fig.~\ref{fig_AEcurrents} (red circles), together with the values estimated from References~\cite{Miseikis_APL100_133105_12} (blue stars) and \cite{Bandhu_APL103_133101_13} (black squares) for the case of transferred graphene. It is remarkable that the three set of data follow a similar linear dependence, which means that the performance of our devices is comparable to the ones based on transferred graphene. We attribute this to the fact that the weaker piezoelectricity of the SiC/MgO/ZnO structure compared to that of the LiNbO$_3$ substrate is compensated in our devices by their better carrier mobility and larger working SAW frequencies. 

\section{Conclusions}\label{sec_Concl}

In this contribution, we have demonstrated the generation of acousto-electric currents in epitaxial graphene on SiC capped with a MgO/ZnO film. The 15 nm-thick MgO layer protects the EG during the sputtering of a thicker ZnO film responsible for the efficient generation of surface acoustic waves. We have demonstrated using Raman and electric characterization that the coating of EG by the MgO/ZnO bilayer does not deteriorate its electronic properties, measuring acousto-electric current densities of the order of $10^{-3}$~A/m. This is one order of magnitude larger than the current densities previously reported in similar devices using HSQ as protective layer. We attribute this enhancement to the larger mobility of the EG coated by MgO with respect to the one coated by HSQ. Our experimental results agree reasonably with the values expected from the classical relaxation model of the acousto-electric interaction. In addition, they are comparable to the current densities reported in graphene transferred to strong piezoelectric substrates. This makes EG on SiC a promising candidate for commercial applications of devices based on the interaction between SAWs and graphene.

\section{Acknowledgement}

The authors acknowledge Manfred Ramsteiner for his suggestions, and Alexander Kuznetsov for discussions. We thank Sander Rauwerdink and Hans-Peter Sch\"{o}nherr for assistance in sample processing. A.H.M. acknowledges financial support by the Deutsche Forschungsgemeinschaft (DFG) within the Priority Programme SPP 1459 Graphene. This publication is part of a project that has received funding from the European Union's Horizon 2020 research and innovation programme under the Marie Sk\l{}odowska-Curie grant agreement No 642688.

\bibliographystyle{unsrt}

\end{document}